# Introduction of Additive Particle Theory for Path Integral Approaches

**Ken-ichi Amano**[a*]

[a] *Faculty of Agriculture, Meijo University, Nagoya, Aichi 468-8502, Japan.*

* Correspondence author: K. Amano (amanok@meijo-u.ac.jp)

**ABSTRACT**

Path integral approaches have been used for boson and fermion systems. The path integral approach has been successful in the many-boson system. However, in the many-fermion system, the path integral approach is not feasible due to the sign problem. In this letter, I introduce additive particle (AP) theory in order to generate an approximation method that avoids the sign problem. The AP theory considers one electron as a string polymer, and virtual particles are added into the system. The AP theory is an approximation, but it is constructed to be able to generate the pair distribution function between free electrons and the density of states of the free electrons at an arbitrary temperature. Hence, when the electrostatic interactions are decreased, the AP theory converges to the free electron system. On the other hand, it deviates from the actual system when the electrostatic interactions are increased. Star polymer approximation and extended star polymer approximation are also introduced.

## 1. Introduction

Recently, Ichii *et al*. have developed frequency modulated atomic force microscopy (FM-AFM) that can measure in liquid metals [1,2]. The FM-AFM can measure the atomic structure of the substrate surface in the liquid metal. Force curve between the AFM tip and the substrate in the liquid metal can be also measured with the FM-AFM. It has been found that the solvophobic attractive force in the liquid metal is stronger than that in water, organic solvents, and ionic liquids. Understanding the interaction in the liquid metal is important, because the knowledge may be used for control of dispersion, aggregation, and self-organization of colloidal particles in the liquid metal. It is important also for realization of liquid metal chemistry (LMC) [3,4]. To support such experimental studies, development of a feasible theory that can treat liquid metal is heavily required.

For calculation of the structure and the interaction in the liquid metal, quantum mechanics should be incorporated into the theory. There are several methods in quantum mechanics, wave mechanics, matrix mechanics, and the path integral approach. In the wave mechanics, Schrödinger equation is used to calculate the wave functions of the electrons. Since it is difficult to treat many electrons for the Schrödinger equation, Kohn–Sham equation of the density functional theory (DFT) has been proposed, and it is now a major approach in quantum mechanics. Orbital-free DFT [5] has also been proposed for speed up of the computation. These approaches are very useful, but the liquid metal is still difficult topic for the approaches, because the liquid metal is many-fermion system with a finite temperature. The matrix mechanics derived by Heisenberg and Feynman's path integral are also famous approaches in quantum mechanics. Here, I write about a problem in the non-time dependent path integral approach. In the path integral approach, permutations among ring polymers are conducted and the number of the permutation is enormously large when the many electron system like the liquid metals are treated. When the number of permutations is even, +1 is multiplied by the corresponding partition function, while when the number of mutations is odd, –1 is multiplied by the corresponding partition function [6-9]. Finally, all of the partition functions are summed, and then the system's partition function is calculated. However, due to the addition and the subtraction, a numerical error occurs. When the numerical error is large, the system's partition function becomes negative, which is absolutely inaccurate. Hence, it is named "minus sign problem", "negative sign problem", or "sign problem". It is still one of the big unsolved problems. Hence, in the present letter, I propose an approximation method, additive particle (AP) theory, to avoid the sign problem. However, I notice that it is just a suggestion. Validity of the AP theory will be judged theoretically and/or numerically in our future study.

In order to treat the liquid metal theoretically, I focused on the path integral approaches for many-boson and many-fermion systems. In Chapter 2-1, the AP theory for a many-boson system consisting of a single species of boson is explained. In Chapter 2-2, the AP theory for a many-fermion system consisting of a single species of fermion is explained, where a calculation result obtained from the previous many-boson system is reutilized. In Chapter 2-3, the AP theory for a many-boson system consisting of two species of bosons is explained with reference to the Chapter 2-1. In Chapter 2-4, the AP theory for a many-fermion system consisting of two species of fermions is explained, where calculational techniques in the previous chapters are reutilized. In Chapter 2-5, an approximation method with relatively lower calculation load is shortly explained. Finally, summary is written in Chapter 3.

### 2-1. Many-boson system consisting of a single species of boson

The first paragraph conveys the introduction of the partition function. The partition function ($Z$) for many-boson system or many-fermion system consisting of a single species is expressed as [6-9]

$$Z_I = \sum_{q \in Q} s(q) \int \cdots \int \exp\left(-\beta\left(U + W_q\right)\right) d\boldsymbol{r}_{11} \cdots d\boldsymbol{r}_{1P}\, d\boldsymbol{r}_{21} \cdots d\boldsymbol{r}_{2P} d\boldsymbol{r}_{N1} \cdots d\boldsymbol{r}_{NP} d\boldsymbol{r}_{\mathrm{n}1} \cdots d\boldsymbol{r}_{\mathrm{n}N'}, \quad (1)$$

where the subscript $I$ is B or F that represent the many-boson and many-fermion systems, respectively. In Eq. (1), the coefficients next to the multiple integrals are neglected, because they are canceled out when expectation values are calculated from related estimators (the coefficient is conveniently set to 1). However, I notify that the neglected coefficients are necessary when you calculate the system free energy from the partition function. In the many-boson system, the sign function $s(q)$ is always +1. In the many-fermion-system, $s(q) = +1$ when the number of permutations is even, $s(q) = -1$ when the number of permutations is odd. The permutation is assigned by $q$ and all of the permutations are included in $Q$. $\beta$ is the inverse temperature and expressed as $1/(k_B T)$, where $T$ is Temperature $T$ and $k_B$ is Boltzmann's constant. $U$ is the total potential energy. $W_q$ is indirectly related to the total kinetic energy, and it itself is the total energy in the ring polymers, shapes of which change depending on the permutation $q$. The total energy in the ring polymers is the sum of the potential energies of the springs between the nodes. The spring constant varies depending on the temperature and the number of the nodes $P$ in the ring polymer. $N$ is the number of boson/fermi particles. The positions of the bose/fermi particles are represented from $\boldsymbol{r}_{11}$ to $\boldsymbol{r}_{NP}$. The subscript n near $d\boldsymbol{r}$ represents the non-boson/fermi particle (Boltzmann particle). The number of non-boson/fermi particles (*e.g.*, nuclei) is represented as $N'$. For example, when an electron/nucleus system with electroneutrality condition is considered and the valency of the nucleus is $v$, $N' = N/v$. For simplicity, I rewrite Eq. (1) as follows

$$Z_I = \sum_{q \in Q} s(q) \int \cdots \int \exp\left(-\beta\left(U + W_q\right)\right) d\boldsymbol{R}_{\mathrm{e}}\, d\boldsymbol{R}_{\mathrm{n}}, \tag{2}$$

where $d\boldsymbol{R}_{\mathrm{e}}$ is the collective representation from $d\boldsymbol{r}_{11}$ to $d\boldsymbol{r}_{NP}$. Although there is the subscript "e" that representing the electron in the case of many-fermion system, "e" represents boson particle in the system of many-boson particles. $d\boldsymbol{R}_{\mathrm{n}}$ is the collective representation from $d\boldsymbol{r}_{\mathrm{n}1}$ to $d\boldsymbol{r}_{\mathrm{n}N'}$.

Our objective here is to explain how to handle the many-*fermion* system so as to avoid the sign problem. However, the AP theory requires a preparation process in the many-boson system beforehand. Hence, the many-boson system is considered here. The partition function that neglecting the coefficients next to the multiple integrals is written as

$$Z_{\mathrm{B}} = \sum_{q \in Q} s(q) \int \cdots \int \exp\left(-\beta\left(U + W_q\right)\right) d\boldsymbol{R}_{\mathrm{e}}\, d\boldsymbol{R}_{\mathrm{n}}\,. \tag{3}$$

The sign function is always +1 and the permutations does not relate to $U$, Eq. (3) can be rewritten as

$$Z_{\mathrm{B}} = \int \cdots \int \exp(-\beta U) \sum_{q \in Q} \exp\left(-\beta W_q\right) d\boldsymbol{R}_{\mathrm{e}} d\boldsymbol{R}_{\mathrm{n}}\,. \tag{4}$$

The ring polymer has the start node and the end node. Only the spring that starts from the end node is changed by the permutation [7-9]. The ring polymer can be seen as the string polymer and the permutable spring. Fortunately, the sum of the energy in the string polymer does not change during the permutation. Then, the partition function can be rewritten as

$$Z_{\mathrm{B}} = \int \cdots \int \exp(-\beta U) \sum_{q \in Q} \exp\big(-\beta(W_{\mathrm{s}} + \tau_q)\big) d\boldsymbol{R}_{\mathrm{e}} d\boldsymbol{R}_{\mathrm{n}} , \tag{5}$$

where $W_{\mathrm{s}}$ is the sum of the energies in the string polymer and $\tau_{\mathrm{q}}$ is the sum of the energies of the permutable springs. Note that $\tau_{\mathrm{q}}$ depends on the permutation $q$. Considering the summation in the integrand, it can be seen as the sum of the Boltzmann factors, and hence it can be viewed as a partition function in a subspace:

$$Z_{\mathrm{BS}} = \sum_{q \in Q} \exp\big(-\beta(W_{\mathrm{s}} + \tau_q)\big) . \tag{6}$$

The subscript S on the left-hand side represents the subspace. Let me introduce a situation that so many boson particles exist in the system almost randomly and the energies of the permutable springs can be alternated by potential energies between additive particles and the end nodes. This situation (approximation) generates a new form of the partition function in the subspace,

$$Z_{\mathrm{BS}} = c_{\mathrm{BS}} \int \cdots \int \exp\left(-\beta(W_{\mathrm{s}} + \sum_{all} \tau_{\mathrm{aI}} + \sum_{all} \tau_{\mathrm{bT}} + \sum_{all} \tau_{\mathrm{ab}})\right) d\boldsymbol{r}_{\mathrm{a1}} \cdots d\boldsymbol{r}_{\mathrm{a}M} d\boldsymbol{r}_{\mathrm{b1}} \cdots d\boldsymbol{r}_{\mathrm{b}M} . \tag{7}$$

In Eq. (7), the subscripts a and b represent the additive particles, and the subscripts I and T respectively represent the initial and terminal nodes of the ring polymers of the bosons. The distance depending functions $\tau_{aI}$, $\tau_{bT}$, and $\tau_{ab}$ are the pair potentials between the substances which the subscripts indicate (I deliberately added only the three pair potentials for this theory). The subscript *all* under the summation means all of the pairs. The numbers of the additive particles a and b are both $M$. Their positions are represented from $\boldsymbol{r}_{a1}$ to $\boldsymbol{r}_{aM}$ and $\boldsymbol{r}_{b1}$ to $\boldsymbol{r}_{bM}$. I assume that $M$ is an arbitrary sufficiently large value. Since the new virtual particles are added and there are new integrals, a new coefficient $c_{BS}$ should be adhered on the multiple integrals. Fortunately, such a coefficient will be canceled out in the calculations of the estimation values and it can be ignored after the derivation. However, for a detailed explanation, the coefficient originating from the AP theory is purposely shown in the equation. The coefficient $c_{BS}$ could be considered as a function of $\boldsymbol{R}_e$ being $c_{BS}(\boldsymbol{R}_e)$. This could be a more rigorous modeling of the partition function in the subspace. However, for simplicity, I expect that the $c_{BS}(\boldsymbol{R}_e)$ can be simplified to $c_{BS}$ of a normal coefficient, because the function resulting from the multiple integrals has the variable $\boldsymbol{R}_e$. In this stage, the shapes of $\tau_{aI}$, $\tau_{bT}$, and $\tau_{ab}$ are not known, but I predict that the shapes of $\tau_{aI}$ and $\tau_{bT}$ are the same, and the three shapes of them must be designed carefully, because the shapes largely change the adsorption result [10].

Eq. (7) contains three unknown functions $\tau_{aI}$, $\tau_{bT}$, and $\tau_{ab}$ which must be solved to clear the approximated partition function in the subspace. If one integrates both sides of Eq. (7) with $d\boldsymbol{R}_e$, the left-hand side becomes the exact form of the partition function of the many-free-boson system being $Z_{Bf}$, because it corresponds to Eq. (4) with $U = 0$. Combining the AP theory formula, it can be written as

$$Z_{\mathrm{Bf}} = c_{\mathrm{BS}} \int \cdots \int \exp\left(-\beta(W_{\mathrm{s}} + \sum_{all} \tau_{\mathrm{aI}} + \sum_{all} \tau_{\mathrm{bT}} + \sum_{all} \tau_{\mathrm{ab}})\right) d\boldsymbol{R}_{\mathrm{ab}} d\boldsymbol{R}_{\mathrm{e}}\,, \tag{8}$$

where $d\boldsymbol{R}_{\mathrm{ab}}$ is the collective representation from $d\boldsymbol{r}_{\mathrm{a}1}$ to $d\boldsymbol{r}_{\mathrm{a}M}$ and from $d\boldsymbol{r}_{\mathrm{b}1}$ to $d\boldsymbol{r}_{\mathrm{b}M}$. The subscript f represents the free state.

The many-free-boson system has been well known system, and the value of $Z_{\mathrm{Bf}}$ at an arbitrary temperature has already been known. The pair distribution function between the free boson particles $g_{\mathrm{Bfee}}(r)$ has been elucidated from the wave mechanics in quantum mechanics [11-13], where $r$ is distance between the centers of the boson particles. Although it has been derived by using the wave mechanics with the grand canonical ensemble, when the number of the boson particles and the system volume are efficiently large, one can ignore the difference between the physical quantities derived from the grand canonical ensemble and the canonical ensemble. Since the free energy, entropy, and energy in the canonical ensemble can be derived from the wave mechanics, the density of states $D_{\mathrm{Bf}}(\varepsilon)$ for the many-free-boson system can be obtained by applying the Salasnich's technique [14], where $\varepsilon$ is not the orbital energy but the system energy. By the way, I introduce a special function $\eta(\varepsilon)$ which is defined as $D(\varepsilon)\mathrm{e}^{-\beta\varepsilon}$. Then, I note that the $g_{\mathrm{Bfee}}(r)$ and $\eta_{\mathrm{Bf}}(\varepsilon)$ are important references for optimization of the pair potentials of $\tau_{\mathrm{aI}}$, $\tau_{\mathrm{bT}}$, and $\tau_{\mathrm{ab}}$. If the simulation system for the AP theory is surrounded by hard walls, the shape of the density distribution of ideal free bosons near the hard wall may also be utilized to optimize the pair potential. For such systems with rigid walls, it is considered preferable to apply a theory based on the canonical ensemble in order to optimize the pair potential associated with the additive particles.

If one performs molecular dynamics (MD) simulation [15,16] or Monte Carlo (MC)

simulation [17,18] obeying the detailed balance in the situation of the right-hand side of Eq. (8), the pair distribution function between the free boson particles in the AP theory $g'_{\rm Bfee}(r)$ and the special function $\eta'_{\rm Bf}(\varepsilon)$ in the AP theory may be obtained. When inputted trial pair potential shapes $\tau_{\rm aI}$ (= $\tau_{\rm bI}$) and $\tau_{\rm ab}$ are appropriate, a situation $g_{\rm Bfee}(r) = g'_{\rm Bfee}(r)$ and $\xi_{\rm Bf}(\varepsilon) \propto \xi'_{\rm Bf}(\varepsilon)$ is realized. Here, the number of the unknown functions is 2 and the number of the available equations is also 2, and hence this is a well-posed optimization problem. After the optimization, several estimated values and functions in the system of many boson particles with $U \neq 0$ (not free) may be calculated by using

$$Z_{\rm B} = c_{\rm BS} \int \cdots \int \exp\left(-\beta\big(U + W_{\rm s} + \sum_{all} \tau_{\rm aI} + \sum_{all} \tau_{\rm bT} + \sum_{all} \tau_{\rm ab}\big)\right) d\boldsymbol{R}_{\rm ab} d\boldsymbol{R}_{\rm e} d\boldsymbol{R}_{\rm n} \tag{9}$$

instead of using Eq. (3). I note that the right-hand side of Eq. (9) with the estimator may support the calculation of the estimated value and function, but the value $Z_{\rm B}$ from the right-hand side of Eq. (9) itself is not meaningful due to absence of the neglected coefficients in Eq. (1).

### 2-2. Many-fermion system consisting of a single species of fermion

Aim of this letter is to make the path integral approach for the many-fermion system of the numerically stable and numerically treatable within a finite time. The partition function in the many-fermion system with a single species of fermion is given by

$$Z_{\rm F} = \int \cdots \int \exp(-\beta U) \sum_{q \in Q} s(q) \exp\big(-\beta(W_{\rm s} + \tau_q)\big) d\boldsymbol{R}_{\rm e} d\boldsymbol{R}_{\rm n} \, . \tag{10}$$

It can be separated as follows

$$Z_{\mathrm{F}} = \int \cdots \int \exp(-\beta U) \left[ \sum_{q \in Q(+)} \exp\left(-\beta(W_{\mathrm{s}} + \tau_q)\right) - \sum_{q \in Q(-)} \exp\left(-\beta(W_{\mathrm{s}} + \tau_q)\right) \right] d\boldsymbol{R}_{\mathrm{e}} d\boldsymbol{R}_{\mathrm{n}}, \quad (11)$$

where $Q(+)$ and $Q(-)$ represent ensembles of the positive permutations and the negative permutations, respectively. It can be rewritten as

$$Z_{\mathrm{F}} = \int \cdots \int \exp(-\beta U) \left[ \sum_{q \in Q} \exp\left(-\beta(W_{\mathrm{s}} + \tau_q)\right) - 2 \sum_{q \in Q(-)} \exp\left(-\beta(W_{\mathrm{s}} + \tau_q)\right) \right] d\boldsymbol{R}_{\mathrm{e}} d\boldsymbol{R}_{\mathrm{n}}. \quad (12)$$

The first summation term is $Z_{\mathrm{BS}}$. Let me name the second summation term as the partition function in the virtual subspace system as follows:

$$Z_{\mathrm{VS}} = \sum_{q \in Q(-)} \exp\left(-\beta(W_{\mathrm{s}} + \tau_q)\right). \quad (13)$$

Using the approximation concept of the AP theory, the partition function can be rewritten as

$$Z_{\mathrm{VS}} = c_{\mathrm{VS}} \int \cdots \int \exp\left( -\beta \left( W_{\mathrm{s}} + \sum_{all} \tau_{\alpha \mathrm{I}} + \sum_{all} \tau_{\beta \mathrm{T}} + \sum_{all} \tau_{\alpha\beta} \right) \right) d\boldsymbol{r}_{\alpha 1} \cdots d\boldsymbol{r}_{\alpha M'} d\boldsymbol{r}_{\beta 1} \cdots d\boldsymbol{r}_{\beta M'}. \quad (14)$$

In Eq. (14), the subscripts α and β are the additive particles which is not the additive particles a and b. The subscripts I and T represent the initial and terminal nodes of the ring polymers of the fermions. The distance depending functions $\tau_{\alpha\mathrm{I}}$, $\tau_{\beta\mathrm{T}}$, and $\tau_{\alpha\beta}$ are the pair potential between the substances which the subscripts indicate. The numbers of the additive particles are both $M'$. The coefficient $c_{\mathrm{VS}}$ could be considered as a function of $\boldsymbol{R}_{\mathrm{e}}$ being $c_{\mathrm{VS}}(\boldsymbol{R}_{\mathrm{e}})$. This could be a more rigorous modeling of the partition function in the subspace. However, for simplicity, I expect that the $c_{\mathrm{VS}}(\boldsymbol{R}_{\mathrm{e}})$ can be simplified to $c_{\mathrm{VS}}$ of a normal coefficient, because the function resulting from the multiple integrals has the variable $\boldsymbol{R}_{\mathrm{e}}$. In this stage, the shapes of $\tau_{\alpha\mathrm{I}}$, $\tau_{\beta\mathrm{T}}$, and $\tau_{\alpha\beta}$ are not known, but I predict that the shapes of $\tau_{\alpha\mathrm{I}}$ and $\tau_{\beta\mathrm{T}}$ are the same ($\tau_{\alpha\mathrm{I}}(r) = \tau_{\beta\mathrm{T}}(r)$ where $r$ represents the distance between the focused substances). The three shapes of them must be designed carefully, because the shapes largely change the adsorption result [10]. Eq. (14) contains the unknown functions, which must also be solved to clear the form of the approximated partition function. The subtraction between the summations in the square brackets in Eq. (12) can be seen as a partition function in a subspace

$$Z_{\mathrm{FS}} = \sum_{q \in Q} \exp\left(-\beta(W_{\mathrm{s}} + \tau_q)\right) - 2 \sum_{q \in Q(-)} \exp\left(-\beta(W_{\mathrm{s}} + \tau_q)\right), \tag{15}$$

simple form of which is

$$Z_{\mathrm{FS}} = Z_{\mathrm{BS}} - 2Z_{\mathrm{VS}}. \tag{16}$$

If one integrates both sides of Eq. (16) with $d\boldsymbol{R}_{\mathrm{e}}$, the left-hand side becomes the exact form of the partition function of the many-free-fermion system being $Z_{\mathrm{Ff}}$, because it corresponds to Eq. (12) with $U = 0$. Combining the AP theory formulas, it can be written

as

$$Z_{\mathrm{Ff}} = c_{\mathrm{BS}} \int \cdots \int \exp\left(-\beta\left(W_{\mathrm{s}} + \sum_{all} \tau_{\mathrm{aI}} + \sum_{all} \tau_{\mathrm{bT}} + \sum_{all} \tau_{\mathrm{ab}}\right)\right) d\boldsymbol{R}_{\mathrm{ab}} d\boldsymbol{R}_{\mathrm{e}}$$

$$-2c_{\mathrm{VS}} \int \cdots \int \exp\left(-\beta\left(W_{\mathrm{s}} + \sum_{all} \tau_{\alpha\mathrm{I}} + \sum_{all} \tau_{\beta\mathrm{T}} + \sum_{all} \tau_{\alpha\beta}\right)\right) d\boldsymbol{R}_{\alpha\beta} d\boldsymbol{R}_{\mathrm{e}}, \quad (17)$$

where $d\boldsymbol{R}_{\alpha\beta}$ is the collective representation from $d\boldsymbol{r}_{\alpha 1}$ to $d\boldsymbol{r}_{\alpha M'}$ and from $d\boldsymbol{r}_{\beta 1}$ to $d\boldsymbol{r}_{\beta M'}$. When the values of $M$ and $M'$ are the same, simply one can unite the integration variables as follows: $d\boldsymbol{R}_{\mathrm{ab}} = d\boldsymbol{R}_{\alpha\beta} \equiv d\boldsymbol{R}_{\mathrm{AB}}$. For example, the $\tau$ functions can be expressed with the subscript AB as follows $\tau_{\mathrm{aI}}(\boldsymbol{r}_{\mathrm{A}k},\boldsymbol{r}_{\mathrm{I}l})$, $\tau_{\mathrm{bT}}(\boldsymbol{r}_{\mathrm{B}k},\boldsymbol{r}_{\mathrm{T}l})$, $\tau_{\mathrm{ab}}(\boldsymbol{r}_{\mathrm{A}k},\boldsymbol{r}_{\mathrm{B}l})$, $\tau_{\alpha\mathrm{I}}(\boldsymbol{r}_{\mathrm{A}k},\boldsymbol{r}_{\mathrm{I}l})$, $\tau_{\beta\mathrm{T}}(\boldsymbol{r}_{\mathrm{B}k},\boldsymbol{r}_{\mathrm{T}l})$, and $\tau_{\alpha\beta}(\boldsymbol{r}_{\mathrm{A}k},\boldsymbol{r}_{\mathrm{B}l})$, where the subscripts $k$ and $l$ are an arbitrary identification number of the node of the ring polymer or an arbitrary identification number of the additive particle.

The pair distribution function between the free fermion particles $g_{\mathrm{Ffee}}(r)$ in the situation of the single species may be elucidated by using the wave mechanics in quantum mechanics [11-13]. Since the free energy, entropy, and energy in the canonical ensemble can be derived from the wave mechanics, the density of states $D_{\mathrm{Ff}}(\varepsilon)$ for the many-free-fermion system can be obtained by applying the Salasnich's technique [14]. We note that $\varepsilon$ is not the orbital energy but the system energy. Then, the special function $\eta_{\mathrm{Ff}}(\varepsilon)$ (= $D_{\mathrm{Ff}}(\varepsilon)e^{-\beta\varepsilon}$) can be obtained. If one performs the MD or MC simulation in the situation of the right-hand side of Eq. (17), the pair distribution function between the free fermion particles in the AP theory $g'_{\mathrm{Ffee}}(r)$ and the special function in the AP theory $\eta'_{\mathrm{Ff}}(\varepsilon)$ may be obtained. Here, I set the estimator for $g'_{\mathrm{Ffee}}(r)$ as $\chi_g$, and $g'_{\mathrm{Ffee}}(r)$ can be calculated from

$$g'_{\mathrm{Ffee}}(r) = \frac{Z_{\mathrm{Bf}}}{Z_{\mathrm{Ff}}}\frac{c_{\mathrm{BS}}}{Z_{\mathrm{Bf}}}\int\cdots\int \chi_g \exp\left(-\beta\left(W_{\mathrm{s}} + \sum_{all}\tau_{\mathrm{aI}} + \sum_{all}\tau_{\mathrm{bT}} + \sum_{all}\tau_{\mathrm{ab}}\right)\right) d\boldsymbol{R}_{\mathrm{AB}} d\boldsymbol{R}_{\mathrm{e}}$$

$$-\frac{2Z_{\mathrm{Vf}}}{Z_{\mathrm{Ff}}}\frac{c_{\mathrm{VS}}}{Z_{\mathrm{Vf}}}\int\cdots\int \chi_g \exp\left(-\beta\left(W_{\mathrm{s}} + \sum_{all}\tau_{\alpha\mathrm{I}} + \sum_{all}\tau_{\beta\mathrm{T}} + \sum_{all}\tau_{\alpha\beta}\right)\right) d\boldsymbol{R}_{\mathrm{AB}} d\boldsymbol{R}_{\mathrm{e}}, \tag{18}$$

where $Z_{\mathrm{Vf}}$ is

$$Z_{\mathrm{Vf}} = c_{\mathrm{VS}}\int\cdots\int \exp\left(-\beta\left(W_{\mathrm{s}} + \sum_{all}\tau_{\alpha\mathrm{I}} + \sum_{all}\tau_{\beta\mathrm{T}} + \sum_{all}\tau_{\alpha\beta}\right)\right) d\boldsymbol{R}_{\mathrm{AB}} d\boldsymbol{R}_{\mathrm{e}}. \tag{19}$$

Eq. (18) is rewritten as

$$g'_{\mathrm{Ffee}}(r) = \frac{Z_{\mathrm{Bf}}}{Z_{\mathrm{Ff}}} g'_{\mathrm{Bfee}}(r) - \frac{2Z_{\mathrm{Vf}}}{Z_{\mathrm{Ff}}} g'_{\mathrm{Vfee}}(r), \tag{20}$$

Taking the far-distance limits in the pair distributions, one obtains

$$Z_{\mathrm{Vf}} = (Z_{\mathrm{Bf}} - Z_{\mathrm{Ff}})/2. \tag{21}$$

Hence, $g'_{\mathrm{Ffee}}(r)$ is calculated by using

$$g'_{\mathrm{Ffee}}(r) = \frac{Z_{\mathrm{Bf}}}{Z_{\mathrm{Ff}}} g'_{\mathrm{Bfee}}(r) - \frac{(Z_{\mathrm{Bf}} - Z_{\mathrm{Ff}})}{Z_{\mathrm{Ff}}} g'_{\mathrm{Vfee}}(r). \tag{22}$$

Focusing on the virtual system and alternating $g'_{\mathrm{Bfee}}(r)$ by $g_{\mathrm{Bfee}}(r)$ and $g'_{\mathrm{Ffee}}(r)$ by $g_{\mathrm{Ffee}}(r)$, it is rewritten as

$$g'_{\mathrm{Vfee}}(r) = \frac{Z_{\mathrm{Bf}}}{Z_{\mathrm{Bf}} - Z_{\mathrm{Ff}}} g_{\mathrm{Bfee}}(r) - \frac{Z_{\mathrm{Ff}}}{Z_{\mathrm{Bf}} - Z_{\mathrm{Ff}}} g_{\mathrm{Ffee}}(r). \tag{23}$$

Since all of the partition functions and the pair distribution functions in the right-hand side of Eq. (23) are known from the wave mechanics, the right-hand side can be prepared. By the way, there are functions derived with the grand canonical ensemble and the canonical ensemble in Eq. (23), but when the number of the particles and the system volume are efficiently large, one can ignore the mixed situation. However, for a system with rigid walls, it is considered preferable to apply the results from a theory based on the canonical ensemble. In the MD or MC simulation, the special function in the virtual system is expressed as follows:

$$\xi'_{\mathrm{Vf}}(\varepsilon) = \frac{Z_{\mathrm{Bf}}}{Z_{\mathrm{Bf}} - Z_{\mathrm{Ff}}} \xi_{\mathrm{Bf}}(\varepsilon) - \frac{Z_{\mathrm{Ff}}}{Z_{\mathrm{Bf}} - Z_{\mathrm{Ff}}} \xi_{\mathrm{Ff}}(\varepsilon). \tag{24}$$

Also in Eq. (24), the right-hand side can be prepared using the wave mechanics. Now, the number of the unknown functions are 2 ($\tau_{\alpha\mathrm{I}}$ ($= \tau_{\beta\mathrm{T}}$) and $\tau_{\alpha\beta}$), and 2 functions (Eq. (23) and Eq. (24)) are available, and therefore one can find $\tau_{\alpha\mathrm{I}}$, $\tau_{\beta\mathrm{T}}$, and $\tau_{\alpha\beta}$ as a well-posed optimization problem. By the way, I note that $g'_{\mathrm{Vfee}}(r) \approx g_{\mathrm{Bfee}}(r)$ and $\xi'_{\mathrm{Vf}}(\varepsilon) \approx \xi_{\mathrm{Bf}}(\varepsilon)$ are realized in the lower temperatures, which may be related to the sign problem.

After the optimization and obtention of the pair potentials, several estimated values and functions in the system of many fermion particles with $U \neq 0$ (not free) may be calculated. However, one encounters a new problem as follows. To obtain an expectation value or function of $X$ in the many-fermion system as $\langle X \rangle_{\mathrm{F}}$, the estimator $\chi$ is applied to the partition function of the AP theory:

$$\langle X\rangle_{\mathrm{F}}=\frac{Z_{\mathrm{B}}}{Z_{\mathrm{F}}}\frac{c_{\mathrm{BS}}}{Z_{\mathrm{B}}}\int\cdots\int\chi\exp\left(-\beta\left(U+W_{\mathrm{s}}+\sum_{all}\tau_{\mathrm{aI}}+\sum_{all}\tau_{\mathrm{bT}}+\sum_{all}\tau_{\mathrm{ab}}\right)\right)d\boldsymbol{R}_{\mathrm{AB}}d\boldsymbol{R}_{\mathrm{e}}d\boldsymbol{R}_{\mathrm{n}}$$
$$-2\frac{Z_{\mathrm{V}}}{Z_{\mathrm{F}}}\frac{c_{\mathrm{VS}}}{Z_{\mathrm{V}}}\int\cdots\int\chi\exp\left(-\beta\left(U+W_{\mathrm{s}}+\sum_{all}\tau_{\mathrm{\alpha I}}+\sum_{all}\tau_{\mathrm{\beta T}}+\sum_{all}\tau_{\mathrm{\alpha\beta}}\right)\right)d\boldsymbol{R}_{\mathrm{AB}}d\boldsymbol{R}_{\mathrm{e}}d\boldsymbol{R}_{\mathrm{n}}. \quad (25)$$

The equation can be simplified as

$$\langle X\rangle_{\mathrm{F}}=\lambda_1\langle X\rangle_{\mathrm{B}}-\lambda_2\langle X\rangle_{\mathrm{\alpha}}, \quad (26)$$

where $\lambda_1 = Z_{\mathrm{B}}/Z_{\mathrm{F}}$ and $\lambda_2 = 2Z_{\mathrm{V}}/Z_{\mathrm{F}}$. Although $\lambda_1$ and $\lambda_2$ are unknown coefficients, their values can be searched as follows. When $X$ is the pair distribution function, taking the far-distance limits to the pair distribution functions one obtains $\lambda_2 = \lambda_1 - 1$. In this letter, a mixture of the electrons and the nuclei of the positive ions (the valency is $\nu$) with electroneutrality condition is considered, and hence the following equation is realized

$$-\nu=\int[\nu\rho_{\mathrm{n}}\langle g_{\mathrm{nn}}(r)\rangle_{\mathrm{F}}-\rho_{\mathrm{e}}\langle g_{\mathrm{ne}}(r)\rangle_{\mathrm{F}}]4\pi r^2dr, \quad (27)$$

where $\rho_{\mathrm{n}}$ and $\rho_{\mathrm{e}}$ are the number density of the positive ion and the electron in the bulk, respectively. The electroneutrality equation can be written with a pair of electron-electron and electron-nucleus instead of the pair of nucleus-nucleus and nucleus-electron. The coefficient $\lambda_1$ in Eq. (27) is optimized in order to realize the equality. Therefore, $\lambda_1$ and $\lambda_2$ become the known coefficients. After that, several estimated values and functions other than the pair distribution functions may be calculated by using the $\tau$ functions, $\lambda_1$, and $\lambda_2$ with the MD or MC simulation.

## 2-3. Many-boson system consisting of two species of bosons

In this chapter, we explain the many-boson system consisting of two species of bosons. It is important for starting the calculation of the many-fermion system consisting of two species of fermions. This chapter explains it with reference to the calculation processes in the previous chapters. The partition function for many-boson system consisting of two species is expressed as [19,20]

$$Z_{\mathrm{B}} = \sum_{q\uparrow \in Q\uparrow} \sum_{q\downarrow \in Q\downarrow} s(q_{\uparrow}) s(q_{\downarrow}) \int \cdots \int \exp\left(-\beta\left(U + W_{q\uparrow} + W_{q\downarrow}\right)\right) d\boldsymbol{R}_{\mathrm{e}\uparrow}\, d\boldsymbol{R}_{\mathrm{e}\downarrow} d\boldsymbol{R}_{\mathrm{n}}, \qquad (28)$$

where $\uparrow$ and $\downarrow$ represent the fictitious up and down bosons, respectively. It is modeled that physical properties of the two bosons are the same except for the identified species up and down. In addition, the number of the up and down bosons are the same. In the above equation, the coefficients next to the multiple integrals are neglected, because they are canceled out when expectation values are calculated from related estimators. It can be rewritten as

$$Z_{\mathrm{B}} = \int \cdots \int \exp(-\beta U) \sum_{q\uparrow \in Q\uparrow} \sum_{q\downarrow \in Q\downarrow} \exp\left(-\beta\left(W_{q\uparrow} + W_{q\downarrow}\right)\right) d\boldsymbol{R}_{\mathrm{e}\uparrow}\, d\boldsymbol{R}_{\mathrm{e}\downarrow} d\boldsymbol{R}_{\mathrm{n}}. \qquad (29)$$

The summation part can be seen as a partition function in a subspace,

$$Z_{\mathrm{BS}} = \sum_{q\uparrow \in Q\uparrow} \sum_{q\downarrow \in Q\downarrow} \exp\left(-\beta\left(W_{q\uparrow} + W_{q\downarrow}\right)\right). \tag{30}$$

In the manner of the AP theory, the ring polymer is modeled as the string polymer and the permutable spring, and hence

$$Z_{\mathrm{BS}} = \sum_{q\uparrow \in Q\uparrow} \sum_{q\downarrow \in Q\downarrow} \exp\left(-\beta\left(W_{\mathrm{s}\uparrow} + \tau_{q\uparrow} + W_{\mathrm{s}\downarrow} + \tau_{q\downarrow}\right)\right), \tag{31}$$

$$Z_{\mathrm{BS}} = \left[\sum_{q\uparrow \in Q\uparrow} \exp\left(-\beta\left(W_{\mathrm{s}\uparrow} + \tau_{q\uparrow}\right)\right)\right]\left[\sum_{q\downarrow \in Q\downarrow} \exp\left(-\beta\left(W_{\mathrm{s}\downarrow} + \tau_{q\downarrow}\right)\right)\right], \tag{32}$$

$$Z_{\mathrm{BS}} = \prod_{j=\uparrow \text{ or } \downarrow} c_{\mathrm{BS}j} \int \cdots \int \exp\left(-\beta\left(W_{\mathrm{s}j} + \sum_{all} \tau_{\mathrm{a}j\mathrm{I}j} + \sum_{all} \tau_{\mathrm{b}j\mathrm{T}j} + \sum_{all} \tau_{\mathrm{a}j\mathrm{b}j}\right)\right) d\boldsymbol{R}_{\mathrm{aj}} d\boldsymbol{R}_{\mathrm{bj}}, \tag{33}$$

$$Z_{\mathrm{BS}} = c'_{\mathrm{BS}} \int \cdots \int \exp\left(-\beta\left(\sum_{j=\uparrow \text{ or } \downarrow}\left(W_{\mathrm{s}j} + \sum_{all} \tau_{\mathrm{a}j\mathrm{I}j} + \sum_{all} \tau_{\mathrm{b}j\mathrm{T}j} + \sum_{all} \tau_{\mathrm{a}j\mathrm{b}j}\right)\right)\right) d\boldsymbol{R}_{\mathrm{ab}\uparrow} d\boldsymbol{R}_{\mathrm{ab}\downarrow}. \tag{34}$$

Here, we note that the coefficient $c_{\mathrm{BS}\uparrow}$ ($c_{\mathrm{BS}\downarrow}$) emerges to connect the summation and the integral, and $c'_{\mathrm{BS}} = c_{\mathrm{BS}\uparrow} c_{\mathrm{BS}\downarrow}$.

Considering the many-free-boson system, the partition function for the system can be written as

$$Z_{\mathrm{Bf}} = c'_{\mathrm{BS}} \int \cdots \int \exp\left( -\beta \left( \sum_{j=\uparrow \text{ or } \downarrow} \left( W_{\mathrm{s}j} + \sum_{all} \tau_{\mathrm{a}j\mathrm{I}j} + \sum_{all} \tau_{\mathrm{b}j\mathrm{T}j} + \sum_{all} \tau_{\mathrm{a}j\mathrm{b}j} \right) \right) \right) d\boldsymbol{R}_{\mathrm{ab}\uparrow\downarrow} d\boldsymbol{R}_{\mathrm{e}\uparrow\downarrow}, \quad (35)$$

where $\mathrm{dR_{ab\uparrow\downarrow}}$ and $\mathrm{dR_{e\uparrow\downarrow}}$ respectively represent $\mathrm{dR_{ab\uparrow}dR_{ab\downarrow}}$ and $\mathrm{dR_{e\uparrow}dR_{e\downarrow}}$. There are six unknown functions in terms τ in the equation, and their variables are different each other. The six unknown functions have different values in most of the cases. However, fortunately, considering the shapes it can be assumed, $\tau_{\mathrm{a\uparrow I\uparrow}}(r) = \tau_{\mathrm{b\uparrow T\uparrow}}(r) = \tau_{\mathrm{a\downarrow I\downarrow}}(r) = \tau_{\mathrm{b\downarrow T\downarrow}}(r)$ and $\tau_{\mathrm{a\uparrow b\uparrow}}(r) = \tau_{\mathrm{a\downarrow b\downarrow}}(r)$, when the numbers of the fictitious up and down bosons are the same. It means that the number of the unknown functions is not six but two. Therefore, the two unknown functions may be inversely calculated by using the $g_{\mathrm{Bfee}}(r)$ and $\xi_{\mathrm{Bf}}(\varepsilon)$ obtained from the wave mechanics in quantum mechanics or a general path integral approach for the binary bosons.

### 2-4. Many-fermion system consisting of two species of fermions

Our final purpose is derivation of the AP theory for the many-fermion system consisting of two species of fermions. The partition function for many-fermion system consisting of two species is expressed as [19,20]

$$Z_{\mathrm{F}} = \sum_{q\uparrow \in Q\uparrow} \sum_{q\downarrow \in Q\downarrow} s(q_{\uparrow}) s(q_{\downarrow}) \int \cdots \int \exp\left( -\beta \left( U + W_{q\uparrow} + W_{q\downarrow} \right) \right) d\boldsymbol{R}_{\mathrm{e}\uparrow}\, d\boldsymbol{R}_{\mathrm{e}\downarrow} d\boldsymbol{R}_{\mathrm{n}}. \quad (36)$$

In the above equation, the coefficients next to the multiple integrals are neglected, because they are canceled out when expectation values are calculated from related estimators. Here, we define the integral part in Eq. (36) as $\varphi$ for simplicity. The partition function for the binary fermions can be rewritten as

$$Z_{\mathrm{F}} = \sum_{q\uparrow\in Q\uparrow(+)} \sum_{q\downarrow\in Q\downarrow(+)} \phi + \sum_{q\uparrow\in Q\uparrow(-)} \sum_{q\downarrow\in Q\downarrow(-)} \phi - \sum_{q\uparrow\in Q\uparrow(+)} \sum_{q\downarrow\in Q\downarrow(-)} \phi - \sum_{q\uparrow\in Q\uparrow(-)} \sum_{q\downarrow\in Q\downarrow(+)} \phi, \quad (37)$$

where $Q\uparrow(+)$ and $Q\uparrow(-)$ represent ensembles of the positive permutations and the negative permutations, respectively. $Q\downarrow(+)$ and $Q\downarrow(-)$ are also the same meanings but only the state is different. The partition function can be simplified as

$$Z_{\mathrm{F}} = \sum_{q\uparrow\in Q\uparrow} \sum_{q\downarrow\in Q\downarrow} \phi - 2 \sum_{q\uparrow\in Q\uparrow(+)} \sum_{q\downarrow\in Q\downarrow(-)} \phi - 2 \sum_{q\uparrow\in Q\uparrow(-)} \sum_{q\downarrow\in Q\downarrow(+)} \phi. \quad (38)$$

The sets of the two summations and $\varphi$ generate the corresponding partition functions, and hence it is given by

$$Z_{\mathrm{F}} = Z_{\mathrm{B}} - 2Z_{\mathrm{V}(+-)} - 2Z_{\mathrm{V}(-+)} = Z_{\mathrm{B}} - 4Z_{\mathrm{V}(\pm)}. \quad (39)$$

Here, I introduced $Z_{\mathrm{V}(+-)} = Z_{\mathrm{V}(-+)} \equiv Z_{\mathrm{V}(\pm)}$, because the numbers of up and down fermions are the same. The partition function in the virtual system is

$$Z_{\mathrm{V}(\pm)} = \sum_{q\uparrow\in Q\uparrow(+)} \sum_{q\downarrow\in Q\downarrow(-)} \int \cdots \int \exp\left(-\beta\left(U + W_{q\uparrow} + W_{q\downarrow}\right)\right) d\boldsymbol{R}_{\mathrm{e}\uparrow\downarrow}\, d\boldsymbol{R}_{\mathrm{n}}. \quad (40)$$

It is rewritten as

$$Z_{\mathrm{V}(\pm)} = \int \cdots \int \exp(-\beta U) \sum_{q\uparrow\in Q\uparrow(+)} \sum_{q\downarrow\in Q\downarrow(-)} \exp\left(-\beta\left(W_{q\uparrow} + W_{q\downarrow}\right)\right) d\boldsymbol{R}_{\mathrm{e}\uparrow\downarrow}\, d\boldsymbol{R}_{\mathrm{n}}. \quad (41)$$

The double summation part can be seen as the partition function of the virtual subspace:

$$Z_{\mathrm{VS}(\pm)} = \sum_{q\uparrow\in Q\uparrow(+)} \sum_{q\downarrow\in Q\downarrow(-)} \exp\left(-\beta\left(W_{q\uparrow} + W_{q\downarrow}\right)\right). \tag{42}$$

In the manner of the AP theory, it is expressed as

$$Z_{\mathrm{VS}(\pm)} = c'_{\mathrm{VS}} \int \cdots \int \exp\left(-\beta\left(\sum_{j=\uparrow\text{ or }\downarrow}\left(W_{\mathrm{s}j} + \sum_{all} \tau_{\alpha j \mathrm{I} j} + \sum_{all} \tau_{\beta j \mathrm{T} j} + \sum_{all} \tau_{\alpha j \beta j}\right)\right)\right) d\boldsymbol{R}_{\alpha\beta\uparrow\downarrow}, \tag{43}$$

where $c'_{\mathrm{VS}} \equiv c_{\mathrm{VS}\uparrow(+)}c_{\mathrm{VS}\downarrow(-)}$. The coefficients $c_{\mathrm{VS}\uparrow(+)}$ and $c_{\mathrm{VS}\downarrow(-)}$ are coproducts which emerged to connect the summation and the integral parts.

The partition function in the virtual system with $U$ (not for the subspace) can be written as

$$Z_{\mathrm{V}(\pm)} = \int \cdots \int \exp(-\beta U) Z_{\mathrm{VS}(\pm)}\, d\boldsymbol{R}_{\mathrm{e}\uparrow\downarrow} d\boldsymbol{R}_{\mathrm{n}}. \tag{44}$$

When $U = 0$ of the many-free-fermion system, a following equation can be written

$$Z_{\mathrm{Ff}} = Z_{\mathrm{Bf}} - 4Z_{\mathrm{V}(\pm)f}, \tag{45}$$

where the subscript f represents the free state. The AP theory expresses $Z_{\mathrm{Ff}}$ as follows:

$$Z_{\mathrm{Ff}} = c'_{\mathrm{BS}} \int \cdots \int \exp\left( -\beta \left( \sum_{j=\uparrow \text{ or } \downarrow} \left( W_{\mathrm{s}j} + \sum_{all} \tau_{\mathrm{a}j\mathrm{I}j} + \sum_{all} \tau_{\mathrm{b}j\mathrm{T}j} + \sum_{all} \tau_{\mathrm{a}j\mathrm{b}j} \right) \right) \right) d\boldsymbol{R}_{\mathrm{ab}\uparrow\downarrow} d\boldsymbol{R}_{\mathrm{e}\uparrow\downarrow}$$

$$- 4c'_{\mathrm{VS}} \int \cdots \int \exp\left( -\beta \left( \sum_{j=\uparrow \text{ or } \downarrow} \left( W_{\mathrm{s}j} + \sum_{all} \tau_{\alpha j\mathrm{I}j} + \sum_{all} \tau_{\beta j\mathrm{T}j} + \sum_{all} \tau_{\alpha j\beta j} \right) \right) \right) d\boldsymbol{R}_{\alpha\beta\uparrow\downarrow} d\boldsymbol{R}_{\mathrm{e}\uparrow\downarrow}. \quad (46)$$

Eq. (46) for the binary system is slightly different from Eq. (17) for the single system. However, the final equational forms of the pair distribution function and the density of states for the virtual system are almost the same as Eq. (23) and Eq. (24), respectively. It is one of the interesting results in the AP theory. There are six unknown functions in terms $\tau$ in the virtual system in Eq. (46) (see the second term). Their distance variables are different from each other. That is, the six unknown functions have different values in most of the cases. However, fortunately, considering the shapes it can be assumed, $\tau_{\alpha\uparrow\mathrm{I}\uparrow}(r) = \tau_{\beta\uparrow\mathrm{T}\uparrow}(r) = \tau_{\alpha\downarrow\mathrm{I}\downarrow}(r) = \tau_{\beta\downarrow\mathrm{T}\downarrow}(r)$ and $\tau_{\alpha\uparrow\beta\uparrow}(r) = \tau_{\alpha\downarrow\beta\downarrow}(r)$, when the numbers of the up and down fermions are the same. It means that the number of the unknown functions is not six but two. Therefore, the two unknown functions may be inversely calculated by using the $g_{\mathrm{Ffee}}(r)$ and $\xi_{\mathrm{Ff}}(\varepsilon)$ obtained from the wave mechanics in quantum mechanics [11-13].

Finally, calculation process of an arbitrary expectation value in the many-fermion-system with the up and down species is explained. When the numbers of the additive particles are the same, simply one can unite the integration variables as follows: $d\boldsymbol{R}_{\mathrm{ab}\uparrow\downarrow} = d\boldsymbol{R}_{\alpha\beta\uparrow\downarrow} \equiv d\boldsymbol{R}_{\mathrm{AB}\uparrow\downarrow}$. For example, the $\tau$ functions can be expressed with the subscripts A and B as follows $\tau_{\mathrm{a}\uparrow\mathrm{I}\uparrow}(\boldsymbol{r}_{\mathrm{A}\uparrow k},\boldsymbol{r}_{\mathrm{I}\uparrow l})$, $\tau_{\mathrm{b}\uparrow\mathrm{T}\uparrow}(\boldsymbol{r}_{\mathrm{B}\uparrow k},\boldsymbol{r}_{\mathrm{T}\uparrow l})$, $\tau_{\mathrm{a}\uparrow\mathrm{b}\uparrow}(\boldsymbol{r}_{\mathrm{A}\uparrow k},\boldsymbol{r}_{\mathrm{B}\uparrow l})$, $\tau_{\alpha\uparrow\mathrm{I}\uparrow}(\boldsymbol{r}_{\mathrm{A}\uparrow k},\boldsymbol{r}_{\mathrm{I}\uparrow l})$, $\tau_{\beta\uparrow\mathrm{T}\uparrow}(\boldsymbol{r}_{\mathrm{B}\uparrow k},\boldsymbol{r}_{\mathrm{T}\uparrow l})$, and $\tau_{\alpha\uparrow\beta\uparrow}(\boldsymbol{r}_{\mathrm{A}\uparrow k},\boldsymbol{r}_{\mathrm{B}\uparrow l})$, where the subscripts $k$ and $l$ are an arbitrary identification number of the node of the ring polymer or an arbitrary identification number of the additive particle. Using the above description method, an arbitrary estimated value and a function in the system of many up and down fermions with $U \neq 0$ (not free) may be calculated with a corresponding estimator $\chi$:

$$\langle X \rangle_{\mathrm{F}} = \theta_{\mathrm{B}} \int \cdots \int \chi \exp\left( -\beta \left( U + \sum_{j=\uparrow \text{ or } \downarrow} \left( W_{\mathrm{s}j} + \sum_{all} \tau_{\mathrm{a}j\mathrm{I}j} + \sum_{all} \tau_{\mathrm{b}j\mathrm{T}j} + \sum_{all} \tau_{\mathrm{a}j\mathrm{b}j} \right) \right) \right) d\boldsymbol{R}'$$

$$- \theta_{\mathrm{V}} \int \cdots \int \chi \exp\left( -\beta \left( U + \sum_{j=\uparrow \text{ or } \downarrow} \left( W_{\mathrm{s}j} + \sum_{all} \tau_{\alpha j\mathrm{I}j} + \sum_{all} \tau_{\beta j\mathrm{T}j} + \sum_{all} \tau_{\alpha j\beta j} \right) \right) \right) d\boldsymbol{R}'. \quad (47)$$

The coefficients $\theta_{\mathrm{B}}$ and $\theta_{\mathrm{V}}$ are expressed as

$$\theta_{\mathrm{B}} = \frac{Z_{\mathrm{B}} c'_{\mathrm{BS}}}{Z_{\mathrm{F}} Z_{\mathrm{B}}} \quad (48)$$

and

$$\theta_{\mathrm{V}} = \frac{4 Z_{\mathrm{V}(\pm)} c'_{\mathrm{VS}}}{Z_{\mathrm{F}} Z_{\mathrm{V}(\pm)}}, \quad (49)$$

respectively. We intentionally placed the same partition functions in both the numerator and the denominator in the above equations. The integral variable $d\boldsymbol{R}'$ means $d\boldsymbol{R}_{\mathrm{AB}\uparrow\downarrow} d\boldsymbol{R}_{\mathrm{e}\uparrow\downarrow} d\boldsymbol{R}_{\mathrm{n}}$. Eq. (47) can be simplified as

$$\langle X \rangle_{\mathrm{F}} = \lambda'_1 \langle X \rangle_{\mathrm{B}} - \lambda'_2 \langle X \rangle_{\mathrm{V}}, \quad (50)$$

where $\lambda'_1 = Z_{\mathrm{B}}/Z_{\mathrm{F}}$ and $\lambda'_2 = 4Z_{\mathrm{V}(\pm)}/Z_{\mathrm{F}}$. Fortunately, we could obtain almost the same form of Eq. (26) also in the present binary system. Hence, the expectation value may be obtained through the process written below Eq. (26).

## 2-5. Star Polymer Approximation and Extended Star Polymer Approximation

Calculation load of the AP theory may be low in comparison with the normal path integral approach. However, the AP theory still has a high calculation load, because it requires the inverse calculation of multiple pair potentials related to the additive particles. Here, I introduce star polymer (SP) approximation in order to reduce the calculation load by starting from the AP theory. I note that the SP approximation can be also derived from the normal partition function of the path integral approach by using forward and inverse of Taylor expansions and applying probability density functions to the permutable springs. It means that the SP approximation is not a method with large logical leap. I found that the SP approximation is closely related to the first-order cumulant expansion, and hence I derived that of the second-order cumulant expansion. I mention the approximation with such a higher-order cumulant expansion as extended star polymer (ESP) approximation.

To convey results of the SP approximation simply, I show only the resulting formulas for the many-free-boson system with two species (↑ and ↓) and the many-free-fermion system with two species (↑ and ↓). The numbers of the species ↑ and ↓ are the same. Normally, the boson system does not consist of ↑ and ↓, but the two species are introduced to enable calculation of the fermion system continuously. In the AP theory, the additive particles are assumed to move freely within the system. However, reducing this degree of freedom can significantly decrease the computational cost. Accordingly, we assume that the added particles are allowed to move only along the line segments of the permutable springs. Furthermore, we assume that one pair of additive particles exists within each permutable spring. Under these conditions, as can be understood from the partition function presented in our previous optical tweezers study [21], the line integral over the additive particles in the partition function of the additive particles leads to a reduced form that depends solely on the length of permutable spring $\gamma$. By applying this result, the partition function for the many-free-boson system with two species can be written as

$$Z_{\mathrm{Bf}} = N!^2 \int \cdots \int \exp\left( -\beta \sum_{j=\uparrow \text{ or } \downarrow} \left( W_{\mathrm{s}j}(\boldsymbol{R}_{\mathrm{e}j}) + \sum_{k=1}^{N} \sum_{l=1}^{N} f\left( \gamma_{jkl}(\boldsymbol{R}_{\mathrm{e}j}) \right) / N \right) \right) d\boldsymbol{R}_{\mathrm{e}\uparrow\downarrow}, \tag{51}$$

where $f$ is quasi free energy of a permutable spring and $\gamma_{jkl}$ represents the length of the permutable spring of the species $j$ between $k_{\mathrm{th}}$ terminal and $l_{\mathrm{th}}$ terminal nodes. As noted above, Eq. (51) can be derived also from the normal partition function of the path integral approach. Deriving it from the normal start, it was found that $c'_{\mathrm{BS}}{}^{1/2} = c_{\mathrm{BS}\uparrow} = c_{\mathrm{BS}\downarrow} = N!$. Since there is one unknown function $f(\gamma)$ in Eq. (51), it must be inversely calculated. It may be inversely calculated by cross-referencing $g_{\mathrm{Bfee}}(r)$ and $g'_{\mathrm{Bfee}}(r)$. Hence, the SP approximation does not require the cross reference of the density of the state functions. Similarly, the partition function for the many-free-fermion system with two species can be written as

$$Z_{\mathrm{Ff}} = Z_{\mathrm{Bf}} - N!^2 \int \cdots \int \exp\left( -\beta \sum_{j=\uparrow \text{ or } \downarrow} \left( W_{\mathrm{s}j}(\boldsymbol{R}_{\mathrm{e}j}) + \sum_{k=1}^{N} \sum_{l=1}^{N} f_{(\pm)}\left( \gamma_{jkl}(\boldsymbol{R}_{\mathrm{e}j}) \right) / N \right) \right) d\boldsymbol{R}_{\mathrm{e}\uparrow\downarrow}, \tag{52}$$

where $f_{(\pm)}$ is quasi free energy of a permutable spring in the virtual system. Eq. (52) can be derived also from the normal partition function of the path integral approach. Deriving it from the normal start, it was found that $c'_{VS}{}^{1/2} = c_{VS\uparrow(+)} = c_{VS\downarrow(-)} = N!/2$. Since there is one unknown function $f_{(\pm)}(\gamma)$ in Eq. (52), it must be inversely calculated. It may be inversely calculated by cross-referencing $g_{\mathrm{Ffee}}(r)$ and $g'_{\mathrm{Ffee}}(r)$. Hence, the SP approximation does not require the cross reference of the density of the state functions.

As an example of the ESP approximation, I introduce that with the second-order. The detailed derivation process is not written here, but it can be derived applying the probability density function and its symmetry. The partition function for the many-free-boson system with two species can be written as

$$Z_{\mathrm{Bf}} = N!^2 \int \cdots \int \exp\left(-\beta \sum_{j=\uparrow \text{ or } \downarrow} \left(W_{\mathrm{s}j}(\boldsymbol{R}_{\mathrm{e}j}) + \kappa_{1j}(\boldsymbol{R}_{\mathrm{e}j}) - \beta\kappa_{2j}(\boldsymbol{R}_{\mathrm{e}j})/2\right)\right) d\boldsymbol{R}_{\mathrm{e}\uparrow\downarrow}, \tag{53}$$

where $\kappa_{1j}$ and $\kappa_{2j}$ are the first- and second-order cumulant coefficients. $\kappa_{1j}$ can be written as

$$\kappa_{1j}(\boldsymbol{R}_{\mathrm{e}j}) = \langle\tau_{qj}(\boldsymbol{R}_{\mathrm{e}j})\rangle_{Qj} = \frac{1}{N}\sum_{k=1}^{N}\sum_{l=1}^{N} \varepsilon\left(\gamma_{jkl}(\boldsymbol{R}_{\mathrm{e}j})\right), \tag{54}$$

where $\langle X \rangle_{Qj}$ represents the average of $X$ in $Qj$ ($j = \uparrow$ or $\downarrow$) ensemble. $\kappa_{2j}$ can be written as

$$\kappa_{2j}(\boldsymbol{R}_{\mathrm{e}j}) = \langle\tau_{qj}(\boldsymbol{R}_{\mathrm{e}j})^2\rangle_{Qj} - \langle\tau_{qj}(\boldsymbol{R}_{\mathrm{e}j})\rangle_{Qj}{}^2, \tag{55}$$

where $\langle\tau_{qj}(\boldsymbol{R}_{\mathrm{e}j})^2\rangle_{Qj}$ is

$$\langle \tau_{qj}(\boldsymbol{R}_{\mathrm{e}j})^2 \rangle_{Qj} = \frac{1}{N^2 - N + 1}$$

$$\left[ \sum_{k=1}^{N} \sum_{l=1}^{N} \varepsilon\left(\gamma_{jkl}(\boldsymbol{R}_{\mathrm{e}j})\right) \varepsilon\left(\gamma_{jkl}(\boldsymbol{R}_{\mathrm{e}j})\right) + 2 \sum_{k=1}^{N} \sum_{l=1}^{N} \sum_{k\prime=k+1}^{N} \sum_{l\prime=1}^{N} \varepsilon\left(\gamma_{jkl}(\boldsymbol{R}_{\mathrm{e}j})\right) \varepsilon\left(\gamma_{jk\prime l\prime}(\boldsymbol{R}_{\mathrm{e}j})\right) \right]. \quad (56)$$

In Eqs. (54) and (56), $\varepsilon(\gamma)$ is the energy of a permutable spring with length $\gamma$ and $\gamma_{jkl}$ represents the length of the permutable spring of the species $j$ between $k_{\mathrm{th}}$ terminal and $l_{\mathrm{th}}$ terminal nodes. Similar to the SP approximation, $\varepsilon(\gamma)$ can be alternated to that of the quasi free energy $f(\gamma)$. The shape of $f(\gamma)$ is controlled by referring the $g_{\mathrm{Bfee}}(r)$ and $g'_{\mathrm{Bfee}}(r)$.

The ESP theory for the many-free-fermion system with two species can be written as

$$Z_{\mathrm{Ff}} = Z_{\mathrm{Bf}} - N!^2 \int \cdots \int \exp\left( -\beta \sum_{j=\uparrow \text{ or } \downarrow} \left( W_{sj}(\boldsymbol{R}_{\mathrm{e}j}) + \kappa_{1j}'(\boldsymbol{R}_{\mathrm{e}j}) - \beta \kappa_{2j}'(\boldsymbol{R}_{\mathrm{e}j})/2 \right) \right) d\boldsymbol{R}_{\mathrm{e}\uparrow\downarrow}, \quad (57)$$

where $\kappa_{1j}$’ and $\kappa_{2j}$’ are the first- and second-order cumulant coefficients. When the spring energy $\varepsilon(\gamma)$ is alternated to the quasi free energy of a permutable spring in the virtual system $f_{(\pm)}(\gamma)$, $\kappa_{1j}$’ and $\kappa_{2j}$’ are written as follows:

$$\kappa_{1j}'(\boldsymbol{R}_{\mathrm{e}j}) = \langle \tau_{qj}(\boldsymbol{R}_{\mathrm{e}j}) \rangle_{Qj(\pm)} = \frac{1}{N} \sum_{k=1}^{N} \sum_{l=1}^{N} f_{(\pm)}\left(\gamma_{jkl}(\boldsymbol{R}_{\mathrm{e}j})\right); \quad (58)$$

$$\kappa_{2j}'(\boldsymbol{R}_{\mathrm{e}j}) = \langle \tau_{qj}(\boldsymbol{R}_{\mathrm{e}j})^2 \rangle_{Qj(\pm)} - \langle \tau_{qj}(\boldsymbol{R}_{\mathrm{e}j}) \rangle_{Qj(\pm)}^{\ 2}; \quad (59)$$

$$\langle \tau_{qj}(\boldsymbol{R}_{\mathrm{e}j})^2 \rangle_{Qj(\pm)} = \frac{1}{N^2 - N + 1}$$

$$\left[\sum_{k=1}^{N}\sum_{l=1}^{N} f_{(\pm)}\left(\gamma_{jkl}(\boldsymbol{R}_{\mathrm{e}j})\right) f_{(\pm)}\left(\gamma_{jkl}(\boldsymbol{R}_{\mathrm{e}j})\right) + 2\sum_{k=1}^{N}\sum_{l=1}^{N}\sum_{k'=k+1}^{N}\sum_{l'=1}^{N} f_{(\pm)}\left(\gamma_{jkl}(\boldsymbol{R}_{\mathrm{e}j})\right) f_{(\pm)}\left(\gamma_{jk'l'}(\boldsymbol{R}_{\mathrm{e}j})\right)\right]. \quad (60)$$

I note that $<X>_{Qj(\pm)}$ represents the average of $X$ in $Qj_{(+)}$ or $Qj_{(-)}$ ensemble, where $j$ = ↑ or ↓. In the equations, there is an approximation that the ensemble averages are almost the same. By referencing $g_{\mathrm{Ffee}}(r)$ and $g'_{\mathrm{Ffee}}(r)$, shape of $f_{(\pm)}(\gamma)$ may be inversely calculated.

## 3. Summary

In summary, I explained the AP theory where one electron is modeled as a string, and virtual particles are added to the system. First, the virtual pair potential for the additive particle is calculated in the many-free-boson system. Second, the virtual pair potential for the second additive particle is calculated in the many-free-fermion system. Finally, the structure and the interaction in the liquid metal of the many-fermion system is calculated by using the previously obtained two virtual pair potentials. The SP approximation was also explained in order to reduce the calculation load.

The calculation process in the AP theory is briefly explained here: (1) Calculate the pair potentials for the additive particles "a" and "b" in the many-free-boson system of $U = 0$ by using the MD or MC simulation; (2) Calculate the pair potentials for the additive particles "α" and "β" in the virtual system of $U = 0$ by using MD or MC simulation; (3) Calculate the expectation value in the boson system of $U \neq 0$ by using the pair potentials for the additive particles "a" and "b"; (4) Calculate the expectation value in the virtual system of $U \neq 0$ by using the pair potentials for the additive particles "α" and "β"; (5)

Combining the expectation values in the boson and virtual systems, the expectation value in the fermion system of $U \neq 0$ is calculated. By the way, when one used the MD simulation, masses of the additive particles must be prepared. I predict that there will be no large problem as long as the masses are set within the appropriate ranges. This is because what matters is obtaining equilibrium results, not nonequilibrium ones.

This letter is just a suggestion for more stable and less computationally expensive method for the path integral approach. Validity of the AP theory will be theoretically verified or numerically checked by using MC or MD simulation in our future study. It is considered that when the electrostatic interactions are decreased, the approximation can converge to the system of the free electrons. On the other hand, when the electrostatic interactions are increased, the approximation deviate from the actual system. I predict that the AP theory works well at extremely high temperatures. The AP theory is related to not only liquid metals, but also plasma gases [22,23] and conduction electrons in solid metals [24,25], etc. Hence, the verification tests will be performed in such systems.

Finally, I would like to describe an important proposal. I consider that finite-temperature electronic properties, such as electron density distributions, electronic energies, and other related physical properties, can be calculated in a practical manner by employing the following interpolation scheme. As an example, we describe a possible procedure for calculating the electron density distribution at a target temperature. (Step 1): First, the electron density distribution at 0 K is calculated using density functional theory (DFT) or the Hartree-Fock method. (Step 2): If possible, the electron density distribution at one or more finite temperatures is also calculated using a finite-temperature quantum-mechanical method, such as Mermin's finite-temperature DFT. (Step 3): The electron density distribution in the very-high-temperature regime is then calculated using state-of-the-art path-integral methods or other suitable finite-temperature quantum methods. (Step 4): The electron density distribution in the extremely-high-temperature regime is calculated using classical statistical mechanics. (Step 5): Finally, the electron density distribution at the target temperature is obtained by applying an appropriate interpolation function to the results obtained in steps (1)-(4). At present, we believe that the above procedure, consisting of steps (1)-(5), represents a realistic approach for calculating finite-temperature electronic properties at an arbitrary target temperature. From this viewpoint, it is also important that many physicists work together to improve the accuracy of the calculations in steps (1)-(4). Furthermore, it is equally important to develop highly accurate interpolation functions for step (5), either on the basis of theoretical considerations or empirically, for example by utilizing artificial intelligence or other machine-learning techniques. Once finite-temperature electronic properties at an arbitrary target temperature can be calculated using this approach, the resulting data could then be utilized in the development of machine-learning potentials and related methods, making finite-temperature quantum chemical calculations feasible at computational speeds suitable for practical applications.

**Acknowledgements**

I would like to thank R. Iwayasu and H. Kimura for supporting literature searches and theoretical discussions. I would like to convey gratitude for all of the people who supported my study.